\documentclass[twocolumn,floatfix,altaffilletter,superscriptaddress,preprintnumbers,
               tightenlines,showpacs,showkeys,nofootinbib]{revtex4-1}
\pdfoutput=1
\usepackage{epic}
\usepackage{color}
\usepackage{latexsym}
\usepackage{epsf}
\usepackage{epsfig}
\usepackage{amssymb,amsmath}
\usepackage{amsfonts}
\usepackage{dsfont}
\usepackage{graphicx}
\usepackage{epstopdf}
\usepackage{gensymb}
\usepackage{graphics}
\usepackage{subfigure}
\usepackage{multirow}
\usepackage{amssymb} % e.g. \gtrsim
\usepackage{ulem}
\usepackage{amsmath}
\usepackage{bbm} % BlackBoeard letters
\usepackage{mathrsfs}   % e.g. nice Lagrange-L with \mathscr{L}
\usepackage{xspace}
\usepackage{soul}
\usepackage{listings}
\usepackage{siunitx}

\newcommand{\MS}{{\ensuremath{\overline{\text{MS}}}}\xspace}

\usepackage{enumitem}
\usepackage[colorlinks=true,citecolor=blue,linkcolor=blue,urlcolor=blue]{hyperref}
\usepackage[utf8]{inputenc}
\usepackage[capitalise, english]{cleveref}
\crefname{section}{Sec.}{Secs.}
\crefname{table}{Tab.}{Tabs.}

\newcommand{\beq}{\begin{equation}} 
\newcommand{\eeq}{\end{equation}} 
\newcommand{\ba}{\begin{array}}  
\newcommand{\ea}{\end{array}} 
\newcommand{\bea}{\begin{eqnarray}}  
\newcommand{\eea}{\end{eqnarray} }  
\newcommand{\be}{\begin{eqnarray}}  
\newcommand{\ee}{\end{eqnarray} }  
\newcommand{\bal}{\begin{align}}
\newcommand{\eal}{\end{align}}   
\newcommand{\bi}{\begin{itemize}}  
\newcommand{\ei}{\end{itemize}}  
\newcommand{\ben}{\begin{enumerate}}  
\newcommand{\een}{\end{enumerate}}  
\newcommand{\bc}{\begin{center}}
\newcommand{\ec}{\end{center}} 
\newcommand{\bt}{\begin{table}}
\newcommand{\et}{\end{table}}  
\newcommand{\btb}{\begin{tabular}}
\newcommand{\etb}{\end{tabular}}

\newcommand{\eVdist}{\kern-0.06em}
\newcommand{\SARAH}{{\tt SARAH}\xspace}
\newcommand{\Vevacious}{{\tt Vevacious}\xspace}
\newcommand{\SPheno}{{\tt SPheno}\xspace}

\newcommand{\HB}{{\tt HiggsBounds}\xspace}

 \definecolor{mkgreen}{rgb}{0.2,.70,.3}

 \definecolor{fsblue}{rgb}{0.,.0,1.}

\definecolor{tobycolour}{rgb}{.5,.0,.5}

\begin{document}

% --------------------------------------------
\title{The Ultraviolet Landscape of Two-Higgs Doublet Models}
% --------------------------------------------

\author{Manuel E. Krauss}
\email{mkrauss@th.physik.uni-bonn.de} 
\affiliation{Bethe Center for Theoretical Physics \& Physikalisches Institut der 
Universit\"at Bonn,\\ ~~~Nu{\ss}allee 12, D-53115 Bonn, Germany}

\author{Toby Opferkuch}
\email{opferkuch@uni-mainz.de} 
\affiliation{PRISMA Cluster of Excellence and
             Mainz Institute for Theoretical Physics,
             Johannes Gutenberg-Universit\"{a}t Mainz, 55099 Mainz, Germany}

\author{Florian Staub}
\email{florian.staub@kit.edu}
\affiliation{Institute for Theoretical Physics (ITP), Karlsruhe Institute of Technology, Engesserstra{\ss}e 7,\\ ~~~D-76128 Karlsruhe, Germany}
\affiliation{Institute for Nuclear Physics (IKP), Karlsruhe Institute of Technology, Hermann-von-Helmholtz-Platz 1,\\ ~~~D-76344 Eggenstein-Leopoldshafen, Germany}

\preprint{BONN-TH-2018-02, KA-TP-18-2018, MITP/18-062}

% --------------------------------------------
\begin{abstract}
We study the predictions of generic ultraviolet completions of two-Higgs doublet models. We assume that at the matching scale between the two-Higgs doublet model and a ultraviolet complete theory -- which can be anywhere between the TeV and the Planck scale -- arbitrary but perturbative values for the quartic couplings are present. We evaluate the couplings down from the matching scale to the weak scale and study the predictions for the scalar mass spectrum. In particular, we show the importance of radiative corrections which are essential for both an accurate Higgs mass calculation as well as determining the stability of the electroweak vacuum. We study the relation between the mass splitting of the heavy Higgs states and the size of the quartic couplings at the matching scale, finding that only a small class of models exhibit a sizeable mass splitting between the heavy scalars at the weak scale. Moreover, we find a clear correlation between the maximal size of the couplings and the considered matching scale.
\end{abstract}
% --------------------------------------------

\maketitle

% --------------------------------------------
\section{Introduction}
\label{sec:introduction}
% --------------------------------------------

Nowadays, there is hardly any doubt that the particle discovered at the LHC in 2012 with a mass of 125~GeV \cite{Aad:2012tfa,Chatrchyan:2012xdj} is the Higgs boson necessary for electroweak symmetry breaking (EWSB). Although all measured properties of this particle are in good agreement with the predictions of the Standard Model (SM) \cite{deFlorian:2016spz}, it is nevertheless much too early to abandon the possibility that it is only \textit{one} of \textit{several} Higgs scalars at the weak scale. It is therefore crucial to study the properties and predictions of models with extended Higgs sectors. Two-Higgs doublet models (THDMs) are the next-to-minimal extension of the SM Higgs sector, beyond the minimal extension introducing pure gauge singlet scalars. This additional ingredient can be used to study a wide range of effects: deviations in the couplings of the \SI{125}{\GeV} scalar, the presence of additional neutral Higgs scalars (including the possibility of a state lighter than the SM-like one), new effects mediated by charged Higgs bosons, amongst many other new effects not present in the SM. See for instance Ref.~\cite{Branco:2011iw} for a detailed overview of these types of models and their phenomenological implications. On the other hand, THDMs address hardly any of the open questions of the SM. For instance, the hierarchy problem, the nature of dark matter or the mechanism for neutrino masses remain unresolved in minimal THDM realizations. THDMs however, are able to accommodate electroweak baryogenesis, providing new sources of CP-violation as well as a modification of the electroweak phase transition to be first-order \cite{McLerran:1990zh,Turok:1990zg,Cohen:1991iu,Cline:1996mga,Fromme:2006cm}. Modifying the electroweak phase transition requires that one or more of the heavy Higgs masses lie near the SM Higgs mass. However, experimental constraints place lower bounds on the charged Higgs masses, hence a split spectrum implying large quartic couplings is required to realise electroweak baryogenesis \cite{Dorsch:2017nza,Basler:2016obg,Dorsch:2016tab,Dorsch:2014qja,Dorsch:2013wja}. Nevertheless, it is likely that -- if they indeed turn out to be favoured by experiment at some point -- they are only the low-energy limit of a more fundamental theory, such as supersymmetry (SUSY) or a grand unified theories.

Given the large array of possibilities, it is unclear what the ultraviolet (UV) completion of a given THDM might be and at which scale the additional degrees of freedom become relevant. In such a setting, the measurement of a new scalar resonance can shed light on the nature of the UV completion. This expectation arises as THDMs include new renormalisable operators that are therefore unsuppressed by the new physics scale unlike higher dimensional operators induced via new physics. Conversely the absence of any new resonances beyond the SM-like Higgs constrains the space of possible UV completions. There are many studies exploring this avenue via a bottom-up approach, i.e. it is assumed that all properties of a THDM at the weak scale are known and it is checked at which energy scale the theory becomes strongly interacting or suffers from an unstable vacuum \cite{Chakrabarty:2014aya,Chakrabarty:2016smc,Ferreira:2015rha,Chakrabarty:2017qkh,Chowdhury:2015yja,Basler:2017nzu}. Assuming that the fundamental UV theory is weakly interacting at all energies, this then indicates the highest possible scale at which new physics is required. In contrast, there are also studies which use a top-down approach: a specific UV model, usually the simplest realisation of supersymmetry, is assumed and the matching conditions to the THDM are calculated \cite{Haber:1993an,Gorbahn:2009pp,Lee:2015uza}. These couplings are then evolved down to the low scale where one then checks if what is predicted is in agreement with current measurements. However, the minimal supersymmetric Standard Model (MSSM) as a UV completion for THDMs is peculiar as it predicts that the quartic couplings of the THDM at the matching scale are always small because in the MSSM they are necessarily proportional to the square of the gauge couplings. 

Both approaches therefore consider the involved parameters of the theory to be in a very narrow window at the high scale -- either they are so large that a perturbative treatment cannot be trusted any more after this point, or they obey special relations, relegating the quartic couplings to comparatively tiny values. A generic UV completion might, however, look very different in the sense that the Lagrangian parameters can take a much larger variety of values. Examples include non-minimal supersymmetric models like the next-to-minimal supersymmetric SM or composite Higgs models, see e.g. Ref.~\cite{Mrazek:2011iu,Zarate:2016jch}.

In this {work}, we {utilise} a top-down approach, but generalise it to {a diverse array of} UV completions. Hence, we do not make any assumption about the fundamental theory, but allow for arbitrary couplings at the matching scale. The only requirements on the couplings {is that they satisfy} perturbativity and perturbative unitarity. {To obtain} reliable predictions for {weak scale physics}, we perform a state-of-the-art analysis using two-loop renormalisation group equations (RGEs) and a two-loop calculation of the scalar masses. Moreover, the stability of the electroweak vacuum is checked at the one-loop level in contrast to the common approach to rely on tree-level conditions \cite{Barroso:2013awa}. Two-loop RGEs have been applied in earlier works on the high-scale behaviour of THDMs \cite{Chowdhury:2015yja}. However, they were never previously combined with a matching of the couplings at the loop-level. While one {naively expects} that the best approach would be to apply one-loop matching when using two-loop RGEs, it has recently been pointed out that this is not the case \cite{Braathen:2017jvs}: when performing $N$-loop running of the parameters, $N$-loop matching {is required} to {determine all finite} non-logarithmic contributions correctly. This is particularly important in the presence of large couplings, which one often faces in THDMs. Therefore, we find sizeable deviations in the relations between the low- and the high-scale compared to previous studies which only applied a tree-level matching in the bottom-up approach \cite{Cheon:2012rh,Chakrabarty:2014aya,Chakrabarty:2016smc,Ferreira:2015rha,Chakrabarty:2017qkh,Chowdhury:2015yja,Gori:2017qwg,Basler:2017nzu,Dev:2014yca,Das:2015mwa}. This difference is especially pronounced when comparing individual parameter points instead of averaging over the properties of a large set of points.  

This paper is organised at follows: in \cref{sec:model} we fix our conventions for the THDM and define our Ansatz to parametrise the high scale theory. In \cref{sec:results} we discuss the results, pointing out differences and shortcomings of previous approaches, before we conclude in \cref{sec:summary}. In the appendix, we provide details about the calculation of the mass spectrum at loop level.

% --------------------------------------------
\section{The model and the procedure}
\label{sec:model}
% --------------------------------------------

\subsection{The CP-conserving THDM}
% --------------------------------------------

The scalar potential of the CP-conserving THDM reads 
\begin{align}
V &= m_1^2\Phi_1^\dagger \Phi_1 + m_2^2\Phi_2^\dagger \Phi_2 + \lambda_1(\Phi_1^\dagger \Phi_1)^2 + \lambda_2(\Phi_2^\dagger \Phi_2)^2 \notag \\
&\quad+ \lambda_3(\Phi_1^\dagger \Phi_1) (\Phi_2^\dagger \Phi_2) + \lambda_4(\Phi^\dagger_2 \Phi_1)(\Phi^\dagger_1 \Phi_2)  \label{eq:scalar_potential}\\ 
&\quad +  M_{12}^2(\Phi_1^\dagger \Phi_2 + \Phi_2^\dagger \Phi_1) + \frac{\lambda_5}{2}\left( (\Phi_2^\dagger \Phi_1)^2  + (\Phi_1^\dagger \Phi_2)^2 \right)\,. \notag
\end{align}
{Taking} $M_{12}$ and $\lambda_5$ real ensures CP conservation in the scalar sector. Here we have assumed a  $\mathbb Z_2$ symmetry which is softly broken by $M_{12}^2$.\footnote{We assume that the UV completion also respects the $Z_2$ symmetry at least at tree-level, i.e. the additional couplings $\lambda_6 |H_1|^2 (H_1^\dagger H_2)$ and $\lambda_7 |H_2|^2 (H_1^\dagger H_2)$ are at most loop induced like in the MSSM and will be neglected in this study.}  Further note that we have defined all parameters in \cref{eq:scalar_potential} to appear with a positive sign in the potential, i.e. our sign choice for $M_{12}^2$ differs from most definitions in the literature.

After EWSB, the scalar fields can be written as
\begin{align}
\Phi_k = \begin{pmatrix}
\phi^+_k \\ 
\frac{1}{\sqrt{2}} (v_k + \phi^0_k + i\,\sigma_k)
\end{pmatrix}\,,~~~i=1,2\,.
\end{align}
The vacuum expectation values (VEVs) $v_i$ have to fulfil $v_1^2+v_2^2=v^2 \simeq (246\,$GeV$)^2$, and we define their ratio as $\tan\beta = v_2/v_1$. The CP-even neutral scalar fields $\phi^0_i$ mix to form the two mass eigenstates $h$ and $H$ {where} we {will} always denote the SM-like Higgs found at the LHC with $h$. The mixing angle which rotates the gauge into the mass eigenstates is commonly denoted as $\alpha$. The CP-odd states mix to form the physical pseudo-scalar field $A$ as well as the longitudinal component of the $Z$-boson, while the two charged states form a charged Higgs $H^\pm$ and the longitudinal component of the $W$ boson. The pseudo-scalar as well as the charged Higgs mass matrix are diagonalised by a rotation of the angle $\beta$. There are therefore four physical masses and two angles in the scalar sector of the THDM. 
Out of the eight Lagrangian parameters in \cref{eq:scalar_potential}, $m_1^2$ and $m_2^2$ are determined such as to ensure that one is {correctly expanding around the} minimum of the potential which features the correct {pattern of} EWSB.

If the THDM is studied only at the low scale, the quartic couplings can be {treated} as free parameters. Therefore, most of the time in the literature, the five dimensionless parameters $\lambda_i$ are traded for the four masses $m_h,\,m_H,\,m_A$ and $m_{H^\pm}$ as well as the Higgs mixing angle $\alpha$, whereas the soft $\mathbb Z_2$ breaking is directly controlled by choosing $M_{12}^2$. The relations between the physical tree-level observables and the quartic couplings for our conventions {in} \cref{eq:scalar_potential} read \cite{Braathen:2017jvs}
\begin{align}
 \lambda_1 &= \frac{1 + t_\beta^2}{2 (1 + t_\alpha^2) v^2} \left(m_h^2 t_\alpha^2 + m_H^2 + M^2_{12} t_\beta (1+t_\alpha^2) \right) , \label{eq:THDM:couplings_relations_1}\\
 \lambda_2  &= \frac{M_{12}^2(1 + t_\beta^2)}{2 t_\beta^3 v^2} +\frac{(1 + t_\beta^2)\left(m_h^2  + m_H^2 t_\alpha^2 \right)}{2 t_\beta^2 (1 + t_\alpha^2)  v^2}  \,, \label{eq:THDM:couplings_relations_2}\\
 \lambda_3  &= \frac{1}{(1 + t_\alpha^2) t_\beta v^2} \Big[\left(m_H^2-m_h^2\right) t_\alpha (1 + t_\beta^2) \notag\\ 
 &\quad+ 2 m_{H^\pm}^2 (1 + t_\alpha^2) t_\beta + M^2_{12} (1 + t_\alpha^2) (1 + t_\beta^2)\Big]  \,, \label{eq:THDM:couplings_relations_3}\\
 \lambda_4  &=\frac{1}{t_\beta v^2}\left(-M^2_{12}(1+t_\beta^2) + m_A^2 t_\beta - 2 m_{H^\pm}^2 t_\beta\right), \label{eq:THDM:couplings_relations_4} \\
\lambda_5  &= \frac{1}{t_\beta v^2}\left(-M^2_{12}(1+t_\beta^2) - m_A^2 t_\beta \right)\,, \label{eq:THDM:couplings_relations_5}
\end{align}
where $t_x = \tan x$.
The advantage of this translation is obvious: interesting parameter regions can directly be defined by the properties of the spectrum and one doesn't have to {deal with} the Lagrangian parameters {directly}. Of course, one needs to take care that the implicitly assumed Lagrangian parameters are sensible and do not violate unitarity, for instance.\footnote{It has recently been pointed out that for a reliable check of perturbative unitarity in THDMs,  the contributions from finite scattering energies $s$ should {also} be included which were widely ignored before \cite{Goodsell:2018fex}.} However, this is not possible if the THDM is embedded in a more complete framework as we assume here: in that case the quartic couplings are no longer free parameters but are predicted at the matching scale between the THDM and its UV completion -- i.e., there is no direct handle any longer on the masses and mixing angles. Instead, they are predictions at the low scale, to be computed from the running of the couplings while taking care of higher-order corrections.\footnote{These loop corrections necessarily spoil the relations \cref{eq:THDM:couplings_relations_1,eq:THDM:couplings_relations_2,eq:THDM:couplings_relations_3,eq:THDM:couplings_relations_4,eq:THDM:couplings_relations_5} 
which are only valid at tree-level or in an on-shell {renormalisation} scheme. In order to get a connection to the high-scale when working in an on-shell scheme, one needs to calculate the counter-terms $\delta\lambda$ in order to extract the \MS couplings including higher order corrections. These corrected parameters then need to be used in the RGEs when running up in scale. \cite{Braathen:2017jvs}.} This is completely analogous to the approach in studying  constrained versions of SUSY models assuming specific SUSY-breaking mechanisms. 

The Yukawa sector of the model is in principle only a doubling of the SM Yukawa sector in that every one of the two Higgs doublets can couple to quarks and leptons:
\begin{align}
\mathcal L_Y &= - \bar L_L (Y^e_1 \Phi_1 + Y^e_2 \Phi_2)  e_R- \bar Q_L (Y^d_1 \Phi_1 + Y^d_2 \Phi_2)  d_R  \notag \\&\quad+  \bar Q_L (Y^u_1  i \sigma_2 \Phi_1^* + Y^u_2 i \sigma_2 \Phi_2^*)  u_R   + {\rm h.c.}
\end{align}
Here we have suppressed flavour and colour indices. $Q_L$ and $L_L$ are the SM quark and lepton doublets, and $d_R$, $u_R$ and $e_R$ are the right-chiral down- and up-type quarks as well as the right-chiral charged leptons. {The} different types of THDMs are distinguished depending on which Yukawa couplings are {non-zero}. {In what follows we consider only two of the most commonly studied types.} They are defined as:
\begin{itemize}
\item Type-I: fermions only couple to the second Higgs doublet, i.e. $Y_1^a = 0~~\forall ~a=d,u,e$\,,
\item Type-II: down-type fermions couple to $\Phi_1$, up-type fermions to $\Phi_2$, i.e. $Y_1^u = Y_2^d = Y_2^e = 0$\,.
\end{itemize}
Our main results will also hold for the other cases like Type-III or lepto-specific as long as  $\tan\beta$ is small. In this case, the top Yukawa coupling is the only 
large Yukawa coupling {and hence has the largest} impact on the running of the model parameters and the loop corrections. 

\subsection{From the matching scale downwards}
% --------------------------------------------

\begin{figure*}[tb]
\centering
\includegraphics[width=0.45\linewidth]{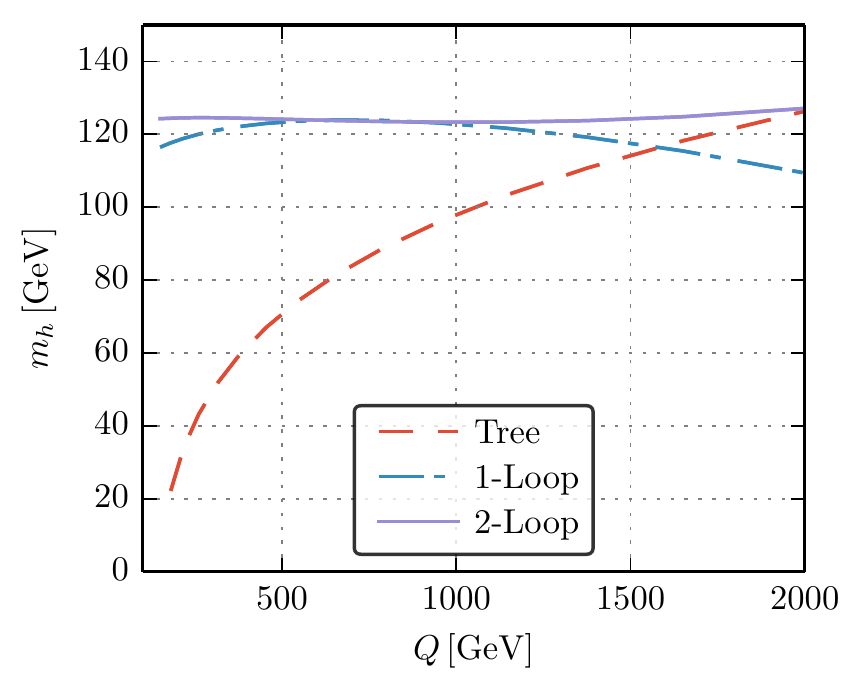} %\hfill
\includegraphics[width=0.45\linewidth]{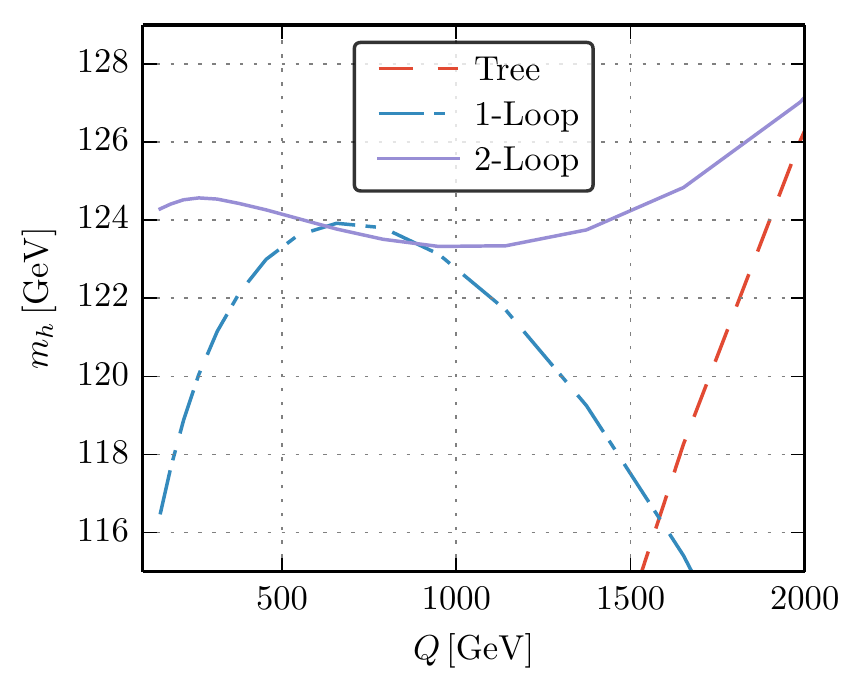}
\caption{The Higgs mass at tree (red), one-loop (blue) and two-loop level (purple) as a function of the renormalisation scale $Q$. As input, we use the running parameters at the top mass scale and evolve them up to $Q$. We have used 
$\lambda_1 = 1.09$, $\lambda_2=0.58$, $\lambda_3=-3.27$, $\lambda_4=0.87$, $\lambda_5=0.81$, $M_{12}=-750^2 \text{GeV}^2$ as well as $\tan\beta=1.18$. {The left-hand pane is a zoomed in version of the right-hand pane, to better illustrate the difference between the 1- and 2-loop computations.}}
\label{fig:Q}
\end{figure*}

Common examples for a UV theory whose low-energy realization is a THDM are for instance high-scale SUSY models with an intermediate $m_A$.
The tree-level matching conditions 
in this case would be \cite{Haber:1984rc} 
\begin{align}
 \lambda_1 &= \lambda_2 = \frac18 \left(g_1^2 + g_2^2\right)\,, & \lambda_3 &= \frac14 \left(g_1^2 - g_2^2\right)\,, \notag\\ 
 \lambda_4 &= -\frac12 g_2^2\,, & \lambda_5 &= 0\,.
\end{align}
It is well known that higher order corrections are important and therefore, the full one-loop as well as dominant two-loop corrections to these matching conditions have been calculated. Still, these corrections don't change the overall magnitude of quartics at the matching scale, i.e. they are still weak couplings with an absolute size smaller than one. This conclusion does, however,  in general not hold for other possible UV completions of the THDM which {can lead to} much larger values for the quartics. In fact, departing from the idea that minimal SUSY must be the fundamental theory behind the THDM, the model parameters could in principle assume any size depending on the details of the UV completion. 
As a concrete example consider a singlet-extended MSSM. In contrast to the most general case, some couplings could be forbidden by an $R$-symmetry for the Higgs and singlet fields. For instance, choosing $R$-charges of 1 for the Higgs doublets, and zero for the singlet, the following superpotential and soft SUSY-breaking terms are allowed:
\begin{align}
W &=\lambda\hat S \hat H_d \hat H_u + \mu \hat H_d \hat H_u + W_Y \\
-\mathcal{L}_{\rm SB} -\mathcal{L}_{{\rm SB},\tilde{f}} &= m_{H_d}^2 |H_d|^2 + m_{H_u}^2 |H_u|^2 + m_S^2 |S|^2 \notag \\
&\quad+  B_S S^2 + T_\kappa S^3 + L S  + \text{c.c.}
\end{align}
where $W_Y$ contains the superpotential terms with Yukawa couplings {as in the MSSM} and $\mathcal{L}_{{\rm SB},\tilde{f}}$ summarises all soft SUSY-breaking terms involving sfermions. If we neglect all contributions from VEVs, the matching conditions at tree level between this model and the THDM become \cite{inprep}
\begin{align}
\lambda_1 = \lambda_2 &= \frac18 \left(g_1^2 + g_2^2 + \frac{4 \sqrt{2} \lambda^2 \mu^2}{M^2} \right)\,, \\
\lambda_3 & = \frac14   \left(-g_1^2+g_2^2 + \frac{12 \sqrt{2} \lambda^2 \mu^2}{M^2}\right) \,,\\
\lambda_4 &= - \frac12 g_2^2 + \lambda^2 \,,\\
\lambda_5 &= \,0 \,,
\end{align}
where $M$ is the mass of the heavy CP-even singlet. Since $\lambda$ is now a free parameter, one can generate much larger values for the quartics of the THDM at tree level. Of course, in this set-up some correlations between the quartics would still exist because they depend on  some fundamental parameters. However, this would also change in even more complicated UV models, especially in non-supersymmetric scenarios where the restrictions on the form of the scalar potential are much weaker.  

Therefore, we are interested in the much more general case and assume that all (perturbative) values of the quartic couplings are allowed at the matching scale $\Lambda$, i.e. they can be in the range
\begin{equation}
 \lambda_i(\Lambda) \in [-4\pi, 4\pi]\,,
\end{equation}
{while also satisfying the} perturbative unitarity {constraints} \cite{Kanemura:1993hm,Akeroyd:2000wc}. Even if we allow in principle for this large range of couplings, we will see that the phenomenologically relevant parameter space is much smaller.

\subsection{Calculating the mass spectrum}
% --------------------------------------------
\begin{table*}[tb]
\begin{tabular}{c|ccccc|cccc}
\hline \hline
RGEs & $\lambda_1(m_t)$   & $\lambda_2(m_t)$ & $\lambda_3(m_t)$ & $\lambda_4(m_t)$ &$\lambda_5(m_t)$&$m_h$[GeV]&$m_{H^0}$\,[GeV]&$m_{A^0}$\,[GeV]&$m_{H^+}$\,[GeV]\\
\hline                
1-loop& 0.304 &  0.202 & -0.168 & -2.331 & 2.067 &  123.6 & 749.1 & 660.4 & 735.8 \\ 
2-loop& 0.370 &  0.243 & -0.084 & -1.948 & 1.695 &  111.6 & 749.4 & 646.0 & 736.2 \\
\hline \hline
\end{tabular}
\caption{The running quartics at the one- and two-loop level for the input given in \cref{eq:exampleRun}. The given values for the masses correspond to a two-loop calculation using $\tan\beta=1.4$ and $M_{12}=-500^2~\text{GeV}^2$}
\label{tab:runLambdas}
\end{table*}

Our goal is to assess the relations between the electroweak (or TeV) scale and a higher scale where the quartic couplings are predicted from matching them to the UV theory.  
As such, it is necessary to treat the couplings as \MS parameters (rather than applying an on-shell scheme) which are then evolved down to the low scale where the spectrum is calculated. In summary, the following steps have to be performed:
\begin{enumerate}
\item Fix the (\MS) couplings at the matching scale $\Lambda$.
\item Evolve the couplings down to the weak scale. For that, we are using the full two-loop RGEs.
\item Calculate the scalar masses and mixing angles including the higher order corrections to the spectrum. In the neutral scalar sector, we compute the full one-loop corrections and add the most important two-loop pieces in the limit of vanishing external momenta. The charged Higgs is calculated at the full one-loop level. 
\end{enumerate}
In \cref{subsec:mass+spectrum}, we provide {additional} details {of} the procedure {used}. The urgent need to go to the two-loop level in order to get a reliable prediction for the Higgs mass in the presence of large 
quartic couplings is demonstrated in \cref{fig:Q} where we show the dependence of the calculated mass on the chosen renormalisation scale. In this example, we use the running quartic couplings as input at the top mass scale and evolve them up to the scale $Q$ where we perform the mass renormalisation. The change in the Higgs mass prediction by varying the renormalisation scale can be used as an {estimate of} the theoretical uncertainty at the different loop-levels. It is seen that, while the scale dependence is huge in the case of a tree-level calculation, the inclusion of the one- and two-loop mass corrections reduces this dependence heavily. Only by including the two-loop corrections we can assume that the theoretical uncertainty is in the ballpark of a few GeV.

\begin{figure}[tb]
\centering
\includegraphics[width=0.9\linewidth]{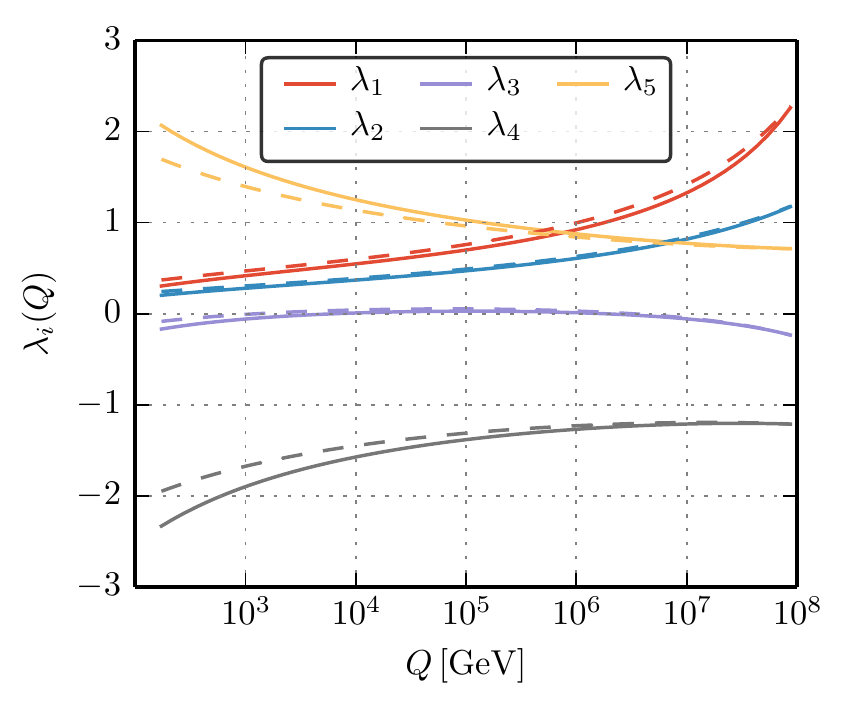} % [width=0.49\linewidth]
\caption{The RGE running at one-loop (dashed) and two-loop (full) of the quartic couplings $\lambda_i$ when fixing the values at the matching scale of $\Lambda=10^8\,$GeV according to \cref{eq:exampleRun}.}
\label{fig:runLambdas}
\end{figure}

Finally we note that the usage of the full two-loop RGEs is, in addition to the radiative Higgs mass corrections, crucial for the accuracy of the predictions when {running from the matching scale down to the top mass scale} since there can be sizeable differences between the one- and two-loop running. In order to demonstrate this, we have chosen the quartic couplings as input at a matching scale of $\Lambda=\SI{E+8}{\GeV}$ to be
\begin{align}
\lambda_1(\Lambda) &= 2.37\,, & \lambda_2(\Lambda) &= 1.21\,, & \lambda_3(\Lambda) &= -0.25\,, \notag\\ 
\lambda_4(\Lambda) &= -1.21\,, & \lambda_5(\Lambda) &= 0.71\,.
\label{eq:exampleRun}
\end{align}
The RGE running at the one- and two-loop level is shown \cref{fig:runLambdas} while the impact on the running quartic couplings as well as the scalar masses is summarised in \cref{tab:runLambdas}.

Here, the change in the light Higgs mass is more than \SI{10}{\GeV}, even with this choice of moderately large quartics at the matching scale. For extreme cases where these couplings approach the limit of $4\pi$, the effects can be much more extreme: points which behave well with two-loop RGEs easily seem to predict tachyonic states at the weak scale if only one-loop RGEs would have been used. 

% --------------------------------------------
\section{Results}
\label{sec:results}
% --------------------------------------------
\subsection{Numerical set-up and constraints}
\label{sec:numerics}
% --------------------------------------------
\subsubsection{Mass spectrum calculation}
\label{subsubsec:massspeccalc}
% --------------------------------------------
For the numerical calculations we make use of the {\tt Mathematica} package \SARAH \cite{Staub:2008uz,Staub:2009bi,Staub:2010jh,Staub:2012pb,Staub:2013tta,Staub:2015kfa} to produce a spectrum generator based on \SPheno \cite{Porod:2003um,Porod:2011nf,Staub:2017jnp}.  As outlined in \cref{subsec:mass+spectrum}, the spectrum is calculated in the \MS scheme at the full one-loop order {including all important two-loop corrections} for the neutral scalars \cite{Goodsell:2014bna,Goodsell:2015ira,Braathen:2017izn}. We have modified the one-loop calculation in such a way that it includes the analytic continuation of loop functions for negative squared masses as input. This is necessary as the one- and two-loop corrections to the SM-like Higgs are often so large that only a negative mass squared at tree-level would lead to a phenomenologically viable spectrum at the two-loop order, otherwise one clearly overshoots the {required} mass of \SI{125}{\GeV}. While it might be uncommon to start with a tachyonic tree-level spectrum, one can think of this as a situation where 
the expansion around the electroweak VEV is a bad one at tree level while the minimum at the right place only emerges at the loop order.\footnote{In some specific supersymmetric models, the only way to obtain a phenomenologically viable spectrum is actually to start with a tachyonic tree-level spectrum which turns into a consistent spectrum at the bottom of a (potentially global) electroweak minimum {appearing} only at the loop level. See for instance Refs.~\cite{Babu:2008ep,Basso:2015pka}.} On the other hand, this issue can be regarded as an artefact of using an \MS scheme. While both the \MS and on-shell scheme {are viable prescriptions to calculate the spectrum}, only the on-shell scheme {enforces the} correct minimum of the potential at every loop order. In \MS, this minimum does not have to be present at every order of perturbation theory, but it has to exist at the highest loop order.

\subsubsection{Scanning procedure}
% --------------------------------------------
 If we would start with random values of the quartics as well as $M_{12}$ at the matching scale and evolve them down, this would correspond to a pure `top-down' approach. However, a parameter scan done in that way would be very inefficient, mainly because  the correct Higgs mass of $m_h \simeq \SI{125}{\GeV}$ would hardly {ever} be obtained. Therefore, we use the more practical Ansatz and scan for $\lambda_i(m_t)$  which give the correct Higgs mass at the two-loop level for given values of $\tan\beta$ and $M_{12}$. These couplings are then evolved up to higher scales using the full two-loop RGEs of the THDM. Here, we are not only interested in the cut-off scale (i.e. the scale where perturbativity or unitarity breaks down) but also all other intermediate scales $\Lambda$. This Ansatz is completely equivalent to  choosing $\lambda_i(\Lambda)$ and $M_{12}(\Lambda)$ randomly at the high scale and keeping only points which have the correct value for $m_h$ and some desired value for $M_{12}$ at the low scale -- with the virtue that we do not have to run the RGEs on points which are being disregarded in the end.

Apart from the quartic couplings, the remaining free model parameters are the soft $\mathbb Z_2$-breaking term $M_{12}^2$ as well as the VEV ratio $\tan\beta$. For $M_{12}^2$, we choose values between $-M_{12}^2 = [0,(\SI{1}{\TeV})^2]$ at the weak scale while {for} $\tan\beta$ {we choose} values between 1 and 2. We have confirmed, by {extending the range of chosen $\tan\beta$ values}, that the $\tan\beta$ dependence of our results is negligible compared to the impact of $\lambda_i$ and $M_{12}$.
\begin{figure*}[tb]
\centering
\includegraphics[width=\linewidth]{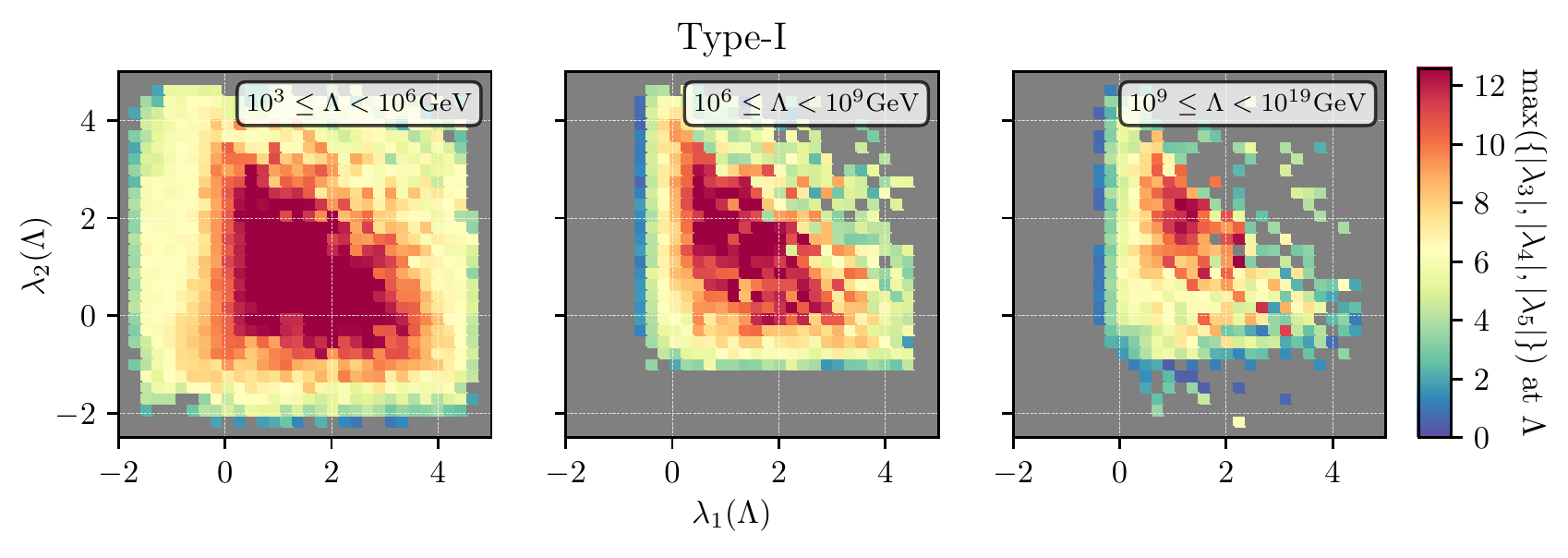}
\caption{The per-bin maximal value of $\{|\lambda_3|,|\lambda_4|,|\lambda_5|\}$ as a function of $\lambda_1$ and $\lambda_2$ at the matching scale. In the plots from left to right, we show three different ranges of the matching scale. We use Yukawa textures of type-I for this figure.}
\label{fig:slam1_vs_lam2}
\end{figure*}

\subsubsection{Theoretical constraints}
\label{subsec:theory_constraints}
% --------------------------------------------
We place several conditions on the {resulting} parameter points. First, as mentioned earlier, we apply the unitarity conditions \cite{Kanemura:1993hm,Akeroyd:2000wc}. For this we {use the quartic couplings entering the two-loop mass spectrum calculation.} Therefore the resulting unitarity constraints, when translated to the physical masses, differ w.r.t. the typical tree-level considerations. Note that this approach is the \MS analogue of using the shifted couplings in an on-shell scheme as proposed in Ref.~\cite{Krauss:2017xpj}.\footnote{These unitarity constraints are obtained in the limit of very large scattering energies, i.e. $\sqrt{s}\to \infty$, and consequently only provide upper limits on combinations of quartic couplings. They should therefore be regarded a very conservative approach. Indeed, it has been shown recently that including the effect of 
trilinear interactions at finite $\sqrt{s}$ can lead to more severe constraints on the parameter space \cite{Goodsell:2018fex,Goodsell:2018tti,Krauss:2018orw}.  While we leave the inclusion of the full results for future work, it has been shown that the expanded perturbativity constraint of demanding that the two-loop scalar mass corrections have to be smaller than the one-loop corrections (which we also apply) leads to quite similar constraints in certain models \cite{Krauss:2018orw}.} We enforce convergence of the perturbative series by demanding that the two-loop correction to all scalar masses has to be smaller than the one-loop corrections, $ |(m_\phi^{2})^{\rm 2L} - (m_\phi^{2})^{ \rm 1L}| < |(m_{\phi}^{2})^{\rm Tree} - (m_\phi^{2})^{\rm 1L}| $, with $\phi=h,H,A$ \cite{Braathen:2017izn,Krauss:2017xpj}. We also apply the conditions for a stable vacuum:
Since loop effects in the Higgs sector are crucial it is not reliable to {use} the common tree-level checks as {has been demonstrated} in Ref.~\cite{Staub:2017ktc}. Instead, we {numerically check} the vacuum stability using the tool \Vevacious \cite{Camargo-Molina:2013qva}. This determines the stability of the one-loop effective potential at the low scale. \Vevacious makes use of the homotopy continuation method provided with {\tt HOM4PS2} \cite{lee2008hom4ps} to find all tree-level extrema of the scalar potential. {It then includes} the one-loop corrections according to Coleman and Weinberg \cite{Coleman:1973jx} and searches numerically for all minima in the vicinity of the tree-level extrema. We only take parameter points into consideration which feature a stable electroweak vacuum, i.e. we disregard regions of parameter space where the electroweak minimum is the false vacuum. The reason is that the tunnelling to minima with VEV values up to a few TeV is very efficient and always leads to a short-lived electroweak vacuum on cosmological time scales.
\footnote{We only consider minima which are `close' to the electroweak one. In this regime, the fixed-order calculation at the one-loop level gives reliable results. For minima involving much larger VEVs, one must consider the RGE-improved potential, also including potentially large effects from gravity. In addition one would need to carefully estimate the tunnelling rate at finite temperature including also the impact of inflation and reheating which was so far done only for the SM \cite{Espinosa:2015qea}.  This is beyond the scope of this paper. Instead, we assume that the vacuum at very high energies can be stabilised by Planck suppressed operators which otherwise do not have any impact on the phenomenological results \cite{Hook:2014uia}.}

\subsubsection{Experimental constraints}
% --------------------------------------------
We also apply the most important experimental constraints. First of all, we demand a SM-like Higgs mass in the range 
\begin{equation}
m_h = \SI{125}{\GeV} \pm \SI{3}{\GeV}\,. 
\end{equation}
This average uncertainty of \SI{3}{\GeV} in the Higgs mass prediction might be too pessimistic for points with small couplings and too optimistic in the presence of huge $\lambda$'s. However, we expect no changes in our results from a model- and parameter point-dependent uncertainty estimate.\footnote{We stick to the fixed estimate because it's not even clear in well established models like the NMSSM how a robust and point-dependent uncertainty estimate should be performed.} 
We furthermore test parameter points against \HB \cite{Bechtle:2008jh,Bechtle:2013wla} to check whether a point is allowed by  Higgs coupling measurements. {Finally}, we impose a lower bound on the charged Higgs mass for the separate cases of type-I and -II Yukawas due to the constraints from $B \to X_s \gamma$ and $B \to X_d \gamma$ \cite{Misiak:2017bgg}. Other flavour constraints, which could be included via the {\tt FlavorKit} functionality of \SPheno \cite{Porod:2014xia} are weaker in the considered scenarios.

\subsection{Numerical Results}
% --------------------------------------------

We now turn to the discussion of the numerical results. We start with summarising the overall results, i.e. what are the preferred values of the quartic couplings at the matching scale, and how does the physics at the weak scale depend on the couplings and the matching scale. Afterwards, we go into detail and analyse the impact of the {included} higher order corrections. 

\subsubsection{The couplings at the matching scale}
% --------------------------------------------
Since $\lambda_1$ and $\lambda_2$ are the most important quartic couplings, i.e. they determine the magnitude of the SM-like Higgs mass as well as (tree-level) vacuum stability, it is natural to {investigate} their possible ranges. Recall in the MSSM $\lambda_1=\lambda_2 >0$ while tree-level vacuum stability of THDMs restricts both $\lambda_1$ and $\lambda_2$ to positive values at the electroweak scale. In \cref{fig:slam1_vs_lam2}, we present the values of $\lambda_1$ and $\lambda_2$ at the matching scale, divided into three ranges of matching scales: \SI{E+3}{}--\SI{E+6}{\GeV} (left), \SI{E+6}{}--\SI{E+9}{\GeV} (middle) and \SI{E+9}{}--\SI{E+19}{\GeV} (right). The maximal (positive) values which we find for these two couplings are {constrained} by perturbative unitarity checks which restrict $\lambda_{1(2)} < 8\pi/6$ if at the same time all other {quartic} couplings {are zero}. 

Thus, even when allowing for a large range of $\lambda_1$ and $\lambda_2$ values at the matching scale, we find that physical constraints {drastically reduce the allowed range of these two quartics}. In particular, large negative values for these couplings {remain} disfavoured by the stability of the electroweak vacuum even when including higher order corrections. The smallest possible value which we found is about $-2$ for low matching scales, while for higher matching scales negative $\lambda_1$ is hardly possible. {Recall that} if one were to apply the tree-level conditions for unbounded-from-below (UFB) directions, one would immediately drop all parameters with negative $\lambda_1$ and/or $\lambda_2$. As a generic result, we also see that large $\lambda_{3,4,5}$ is only allowed at the matching scale if $\lambda_{1,2}$ is moderately large, i.e. $0 \lesssim \lambda_{1,2} \lesssim 3$. The plot on the right-hand side of \cref{fig:slam1_vs_lam2} which displays the case of large matching scales suggests that the larger the matching scale, the smaller the allowed couplings are. This is actually a non-trivial statement {as one} one might have expected that the choice of the matching scale can always be compensated by varying the different quartic couplings without changing the overall magnitude of these couplings. These conclusions remain qualitatively unchanged when looking at the average $\lambda_i$, while the situation in type-II THDMs is very similar.

\subsubsection{The spectrum of THDMs}
% --------------------------------------------
\begin{figure*}[tb]
\centering
\includegraphics[width=\linewidth]{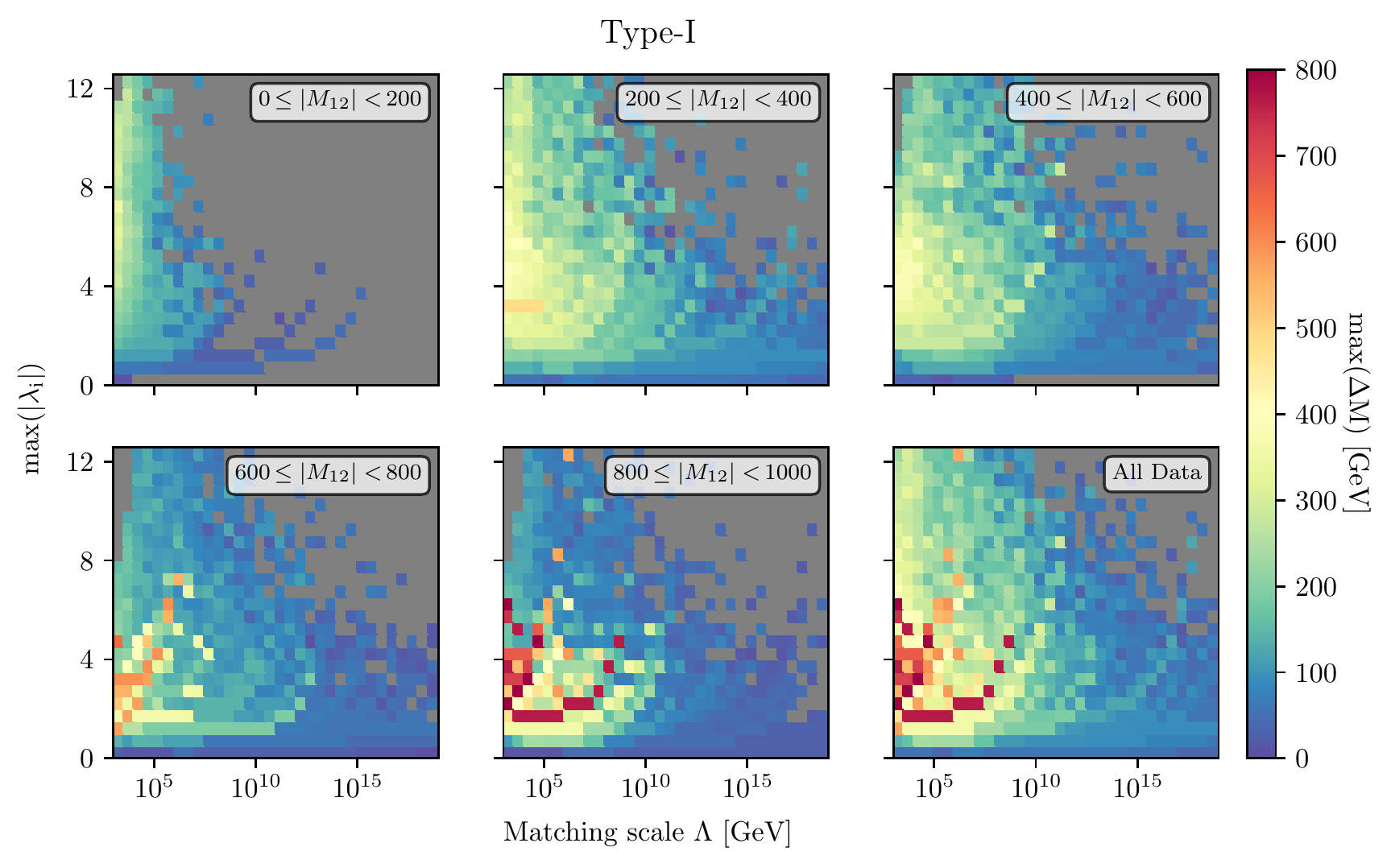}
\caption{The maximal mass splitting between the heavy Higgs states $\Delta M$, evaluated at the electroweak scale, as a function of the matching scale $\Lambda$ and the value of the maximal quartic coupling at $\Lambda$.  We use the Yukawa scheme of type-I for this figure.}
\label{fig:scale_vs_lambda_deltaM}
\end{figure*}
\begin{figure*}[tb]
\centering
\includegraphics[width=\linewidth]{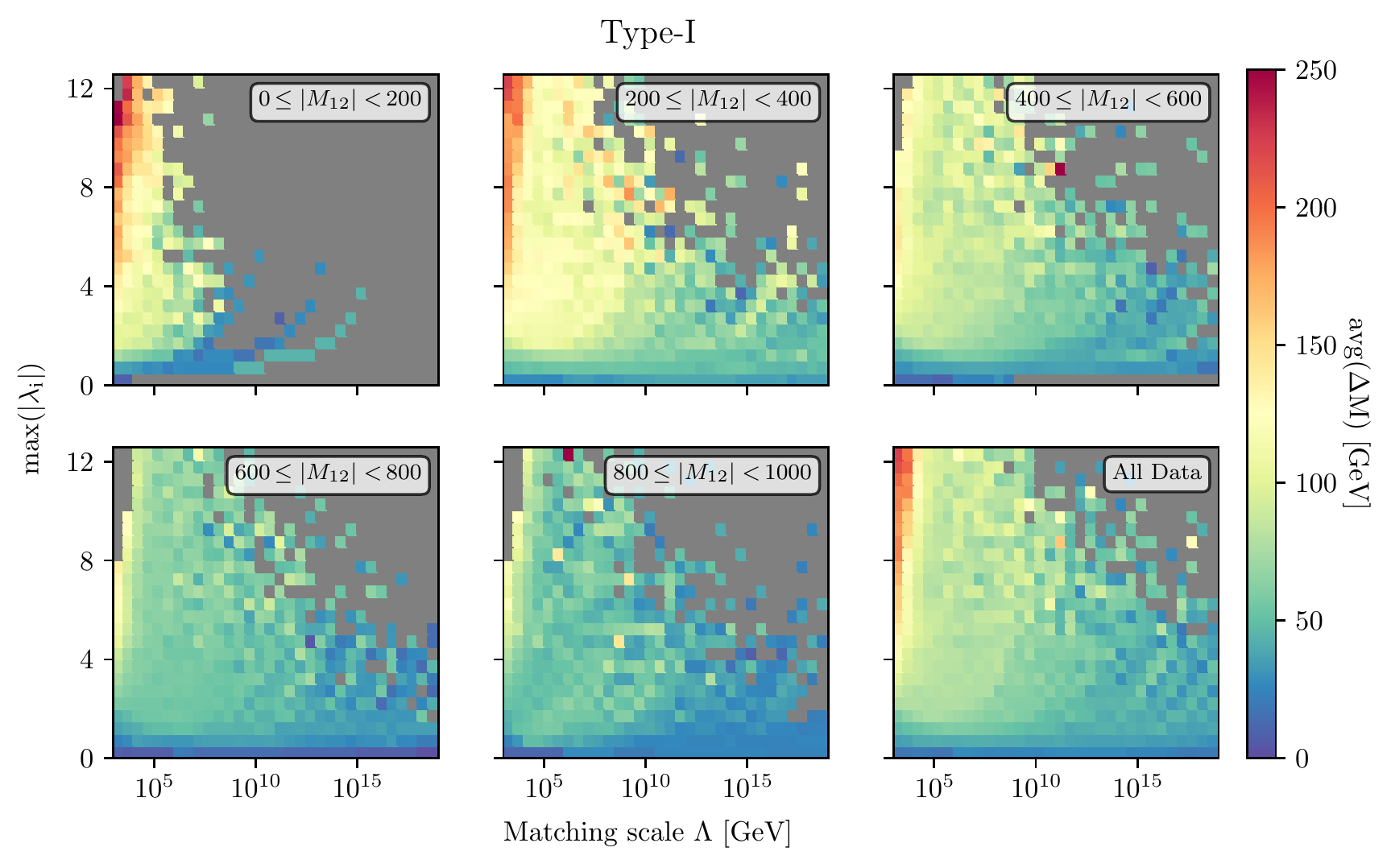}
\caption{Same as \cref{fig:scale_vs_lambda_deltaM} but with the per-bin average of $\Delta M$ instead of the maximal value.}
\label{fig:scale_vs_lambda_deltaM_avg}
\end{figure*}
We turn to the discussion of the scalar mass spectrum. The largest difference between general THDMs and THDMs {arising from a UV completion, like the MSSM,} is that the mass splitting between the heavy scalars can be very large. In contrast, the THDM matched to the MSSM always predicts that the heavy CP-even and -odd Higgs states are nearly degenerate, and the charged Higgs mass only differs by the $W$ boson mass.

We show in \cref{fig:scale_vs_lambda_deltaM} the maximal mass splitting max($\Delta M=|M_i - M_j|$) (evaluated at $m_t$), where $i=H^0, A^0, H^\pm$,  as a function  of the matching scale $\Lambda$ and the maximal, absolute value of the quartic couplings at that scale. We also show the results for different ranges of $M_{12}$. One can draw the connection to previous studies using the bottom-up approach and checking for the cut-off scale of the theory by looking at the region of the plot with the largest quartic couplings, while the MSSM-like parameter region corresponds to the area close to the $x$-axis. In general, we observe that sizeable mass splittings are easier to achieve when matching the THDM to a UV theory at rather low scales. In particular for matching scales above $10^{10}\,$GeV and $|M_{12}|<200$~GeV we rarely find points where the mass splitting between different heavy Higgs states at the electroweak scale turns out to be larger than $\sim 100\,$GeV. If, instead, the mass splitting $\Delta M$ should be of several hundreds of GeV, then this \emph{can not} be realised with matching scales beyond 100\,TeV, particularly so if $M_{12}$ is small. Actually, for this choice of $|M_{12}|$ it is in general difficult to find any valid models at all with large couplings and high matching scale even with small mass splittings in the scalar sector. This statement changes when moving to larger values of $M_{12}$: already for $|M_{12}| \sim 500\,{\rm GeV}$  we can find parameter points which agree with all the electroweak-scale physics while having a matching scale around $10^{10}\,$GeV and mass splittings up to 400\,GeV. For larger values of $|M_{12}|$, this situation does not change significantly as can be seen in the lower row of \cref{fig:scale_vs_lambda_deltaM}.
\begin{figure*}[tb]
\centering
\includegraphics[width=\linewidth]{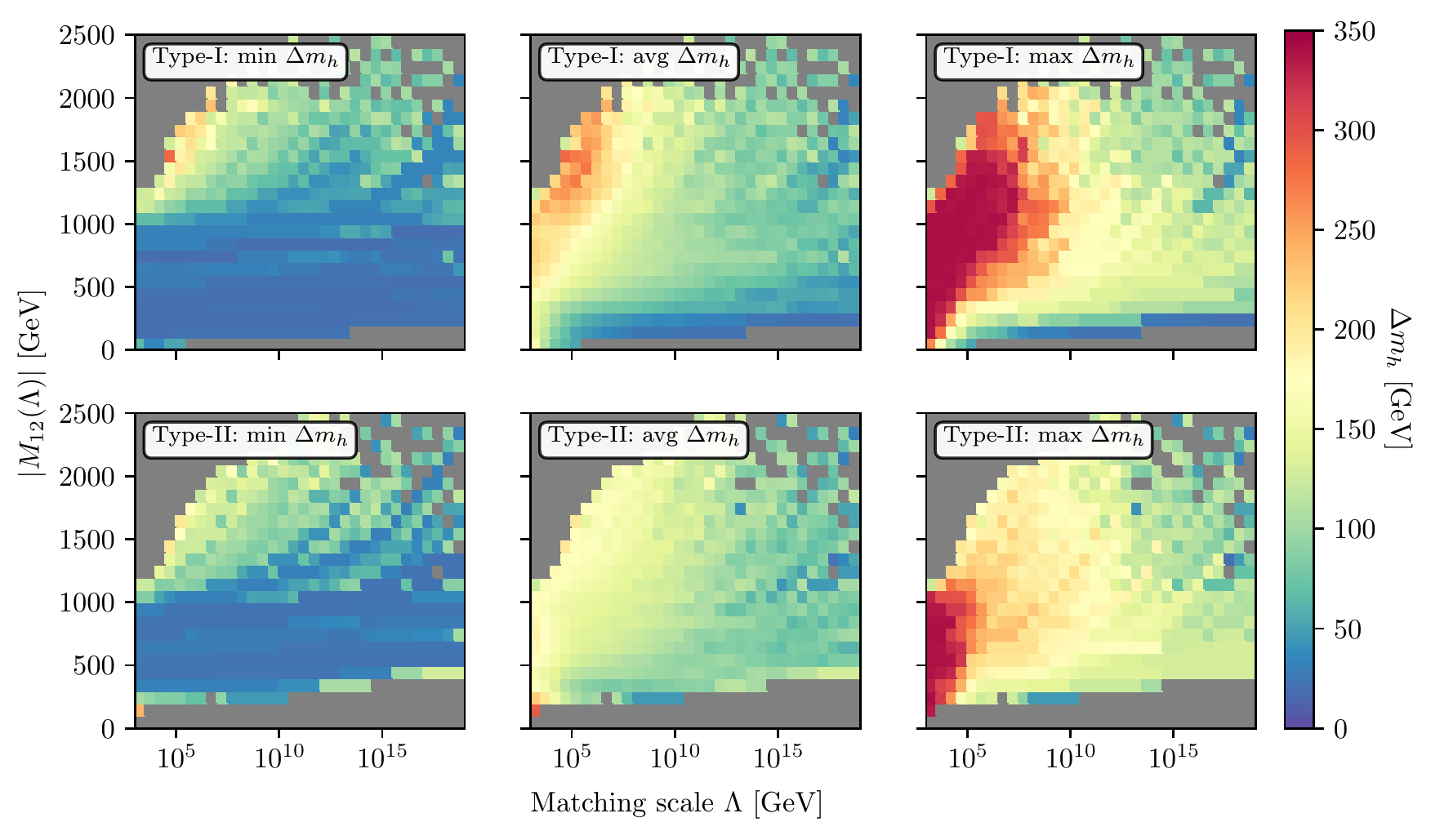} % Cutoff_M12_AllDeltamh
\caption{Size of the loop corrections to $m_h$ as a function of the matching  scale $\Lambda$ and $M_{12}(\Lambda)$. The colours in the plane represent the minimal (left), average (middle) and maximal (right) size of the radiative corrections in the respective bin. We are using type-I Yukawas in the upper row and type-II in the lower row.}
\label{fig:Q_vs_M12_deltamh}
\end{figure*}

The most unexpected feature {is} the largest mass splittings do not appear for the largest values of quartic couplings at the matching scale, but for moderately large couplings of $\mathcal{O}(2-6)$ and large values of $|M_{12}|$ of $\mathcal{O}(\SI{1}{\TeV})$. The reason is that for those value of $M_{12}$ larger quartic couplings are forbidden as the corrections to $m_h$ become too large.

Note that while  \cref{fig:scale_vs_lambda_deltaM} displays the case of type-I Yukawas, the picture is very similar for type-II, and hence the conclusions are the same. The only difference is that the upper left plot featuring $|M_{12}|<200\,$GeV is not populated in the THDM-II because of the {tighter} constraints on the charged Higgs mass from $B$ observables. 

Interestingly, when looking at the average mass splittings between the heavy scalars, we obtain a different picture, as is shown in \cref{{fig:scale_vs_lambda_deltaM_avg}}. {In this figure} we show the per-bin averaged  $\Delta M$ instead of the maximal value {per bin} as before. We find that, while mass splittings of $\sim \SI{150}{\GeV}$ {typically occur} for low $|M_{12}|$, max$(|\lambda_i|) \gtrsim 2$ and matching scales below \SI{E+10}{\GeV}, this is not any more the case for larger values of $|M_{12}|$ where smaller mass splittings of \SI{50}{}--\SI{100}{\GeV} are preferred.

\subsubsection{Impact of scalar loop corrections on the light Higgs}
% --------------------------------------------

We now take  a closer look at the size of the one- and two-loop mass corrections which we obtain for the SM-like Higgs state. As announced earlier, these are generically quite large. Depending on the matching scale, we consider the radiative correction to the Higgs mass (calculated at the top mass scale) as a function of $M_{12}^2$ which we define as
\begin{equation}
\Delta m_h = \sqrt{\left|m_h^{2,\rm loop} - m_h^{2, \rm tree}\right|}\,.
\end{equation}
Here we denote $m_h^{2,{\rm tree}}$ as the mass which we would obtain when calculating the  Higgs mass at tree-level with the quartic couplings that eventually lead to the 125\,GeV at two-loop. It is therefore \emph{not} equivalent to the on-shell mass but can be seen as the input parameter for obtaining the \MS $\lambda_i$ when using \cref{eq:THDM:couplings_relations_1,eq:THDM:couplings_relations_2,eq:THDM:couplings_relations_3,eq:THDM:couplings_relations_4,eq:THDM:couplings_relations_5}. 
 
We show the results for the THDM of both type-I (upper row) and type-II (lower row) in \cref{fig:Q_vs_M12_deltamh}  in the two-dimensional plane matching scale vs $|M_{12}(\Lambda)|$ to represent the minimal, maximal and average Higgs mass correction in the respective bin. 
We see that while the minimal correction is smaller than \SI{100}{\GeV} almost throughout the entire plane, the maximal correction can be as large as \SI{300}{\GeV} for small matching scales and large $|M_{12}|$. The reason for this behaviour is clear: large loop corrections are driven by large couplings at the electroweak scale -- which are more likely to be obtained with low matching scales as can be seen in the previous figures. Even the averaged radiative corrections to the Higgs mass are $\mathcal{O}(\SI{100}{\GeV})$. This shows that a calculation beyond leading order is absolutely crucial for obtaining sensible predictions. Of course, one might feel uncomfortable by these huge loop corrections and wonder about the validity of the perturbative series. As we have stated above, we applied the condition that the two-loop corrections must always be smaller than the one-loop corrections to filter out the most pathological points. In principle, one can apply even stronger constraints on the size of these loop corrections. This would correspond to disfavouring certain classes of UV completions with very large quartic couplings and might be a conservative approach. We always included these extreme parameter regions in order to stress the necessity to include radiative corrections which hasn't been done in literature before.  

The main difference between the cases of type-I and type-II Yukawas {stem from the more stringent constraints on the latter type} \cite{Misiak:2017bgg}, leading to {a lower bound} on $m_{H^\pm_{\rm type-II}}$ of $\mathcal{O}(\SI{600}{\GeV})$. Since $M_{12}$ sets the overall scale of the heavy Higgs states, this cut constrains a combination of $\lambda_i$ and $M_{12}$ and therefore leads to larger minimal $|M_{12}|$ values for type-II models.

\begin{figure*}[tb]
\centering
\includegraphics{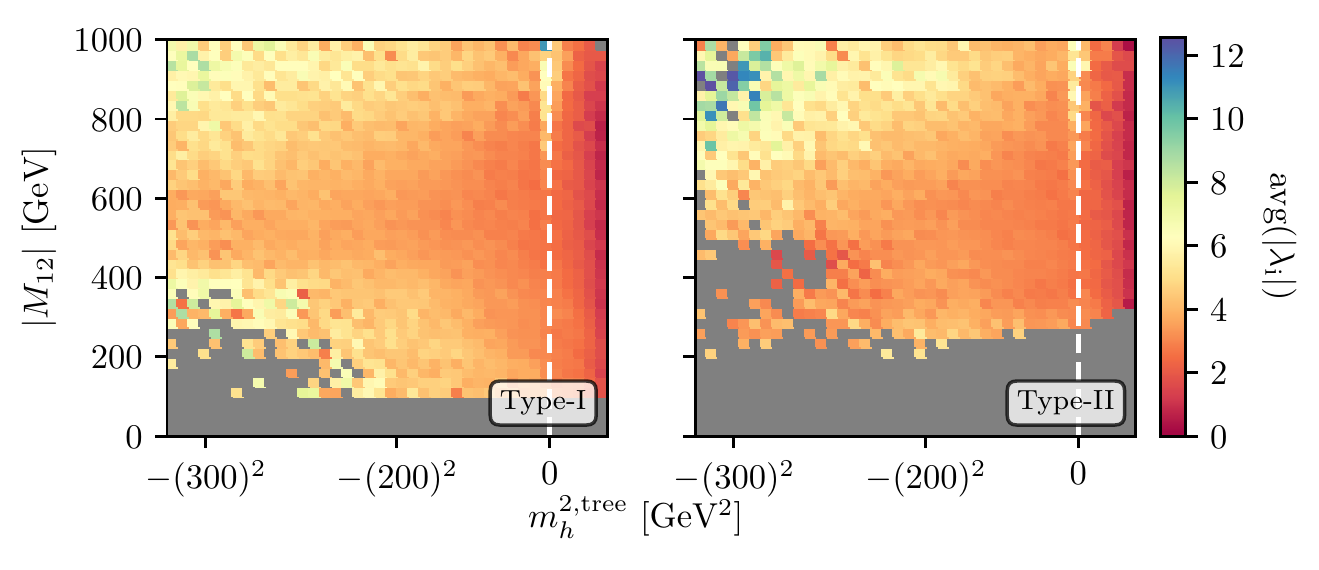} 
\caption{Size of the average electroweak-scale quartic couplings as a function of the tree-level input mass $m_h^{2, {\rm tree}}$ and $|M_{12}|$ using the THDM of type-I (left) and -II (right).}
\label{fig:treemh_m12_avglam}
\end{figure*}

Finally we want to illustrate the ranges of tree-level input parameters that we have to use in order to achieve a 125\,GeV lightest Higgs.
 As explained in \cref{subsubsec:massspeccalc}, it is necessary to often use negative  $m_h^{2,{\rm tree}}$ in order to achieve the correct Higgs mass at the two-loop order. 
In \cref{fig:treemh_m12_avglam}, we present the range  which we used for our study. We contrast this against the electroweak-scale $|M_{12}|$ and show the per-bin average of the $\lambda_i$ in this plane. We observe that valid spectra are only compatible with positive $m_h^{2,{\rm tree}}$ if the quartics are moderate, $\lesssim 4$. {Coupling beyond this value} cause the loop corrections to be so large that negative squared input masses are needed -- and the larger the quartic couplings, the more extreme ranges of $m_h^{2,{\rm tree}}$ are needed. One can see from these large loop corrections that a tree-level study of the Higgs sector is very unreliable. This also underlines the need to test for vacuum stability at the loop level. 

\subsubsection{The sensitivity of the cut-off scale on higher-order corrections}
\label{sec:cutoff}
% --------------------------------------------

The size of the loop corrections discussed in the previous subsection can be translated into the shift in cut-off scale, see Ref.~\cite{Braathen:2017jvs} for further details. We define this scale as the largest scale up to which a perturbative treatment of the THDM is still justifiable, i.e. as the point at which either one of the quartic couplings becomes larger than $4\pi$ or where the perturbative unitarity conditions are not satisfied any more due to the RGE evolution of the $\lambda$'s.
\begin{figure}[tb]
\centering
\includegraphics{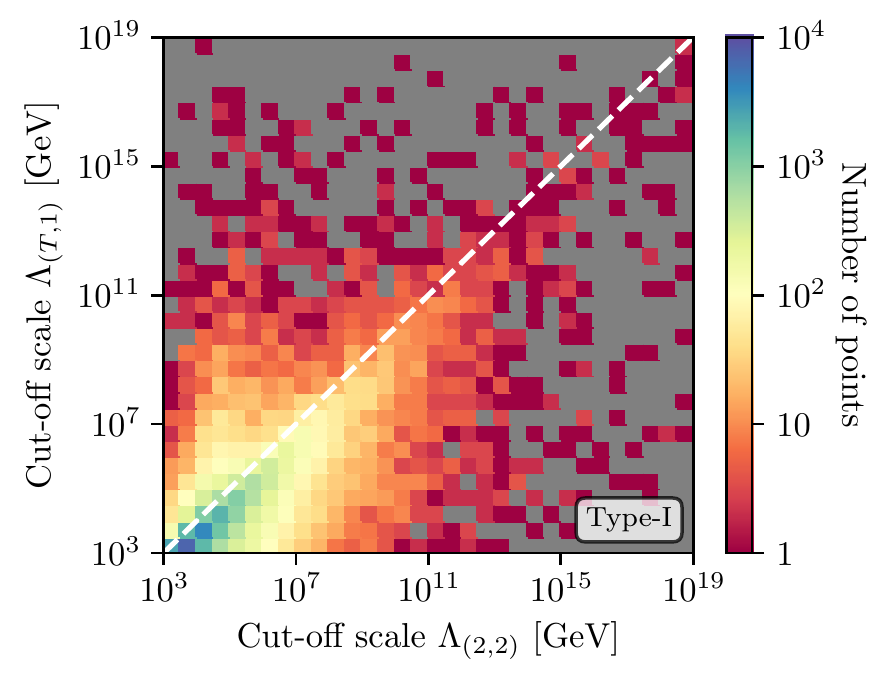}
\caption{Number of parameter points depending on the cut-off scale when including two-loop mass corrections and two-loop RGE running $(2,2)$ as well as when doing a tree-level mass computation and one-loop running $(T,1)$. Here we have applied the Yukawa scheme of type-I. The white line is the diagonal along $\Lambda_{(T,1)} = \Lambda_{(2,2)}$.}
\label{fig:tree_vs_2L_numberofpoints}
\end{figure}

We show {the number of points affected by these considerations in} \cref{fig:tree_vs_2L_numberofpoints}. {More specifically we show t}he cut-off scale when using two-loop Higgs mass corrections at the top mass scale (i.e. a two-loop matching of the THDM to the SM) and two-loop RGE running (denoted ${(2,2)}$) against the cut-off when doing tree-level matching at the weak scale and one-loop running, $(T,1)$. For the latter, we use \cref{eq:THDM:couplings_relations_1,eq:THDM:couplings_relations_2,eq:THDM:couplings_relations_3,eq:THDM:couplings_relations_4,eq:THDM:couplings_relations_5} in order to obtain the tree-level -- or on-shell -- couplings from the mass spectrum for each point.

One first obvious observation is that the majority of points accumulates at low cut-off scales below 100\,TeV -- which is of course no surprise when sampling the parameter space randomly with a flat distribution in the quartics. The second observation is that there is a trend towards higher $(2,2)$ cut-off scales. This is seen as the deviation from the diagonal white line in \cref{fig:tree_vs_2L_numberofpoints} and derives from the fact that the two-loop corrections to the RGEs typically reduce the absolute size of the $\beta$-functions and therefore the slope of the running. Although the majority of the points is characterised by this behaviour, it is interesting to see how drastic the change in matching and running can affect the high-scale behaviour of a parameter region: we find points where the cut-off scale in the $(2,2)$ calculation is larger than the $(T,1)$ prediction by ten orders of magnitude or more -- and vice versa. This is a {direct} consequence of the mostly large one- and two-loop mass corrections to the scalars. The corresponding shift in $\lambda$ when matching at two-loop compared to matching at tree-level can consequently lead to drastic differences in the high-scale behaviour. In particular, it is worth {stressing} that the biggest changes appear for large cut-off scales which correspond to at most moderately large quartic couplings at the weak scale. Thus, for these points the perturbative series at the weak scale behaves well and the loop corrections to the quartics are absolutely trustworthy, while the missing higher-order corrections from three-loop contributions and above can be expected to be small. {As the distribution of points around the white dashed line in \cref{fig:tree_vs_2L_numberofpoints} appears symmetric for larger values of the cut-off, it is tempting to argue that a tree-level mass spectrum calculation in conjunction with one-loop RGEs is sufficient at a statistical level. However, in advocating specific benchmark points, especially for use in experimental searches, it is essential that the complete state-of-the-art calculations be performed to properly ascertain their validity with respect to both theoretical and experimental constraints. } 
{Again} we note that there are no {sizeable} differences between the cases of type-I and type-II Yukawa textures.

\subsubsection{Vacuum Stability}
% --------------------------------------------
\begin{figure*}[t]
\centering
\includegraphics{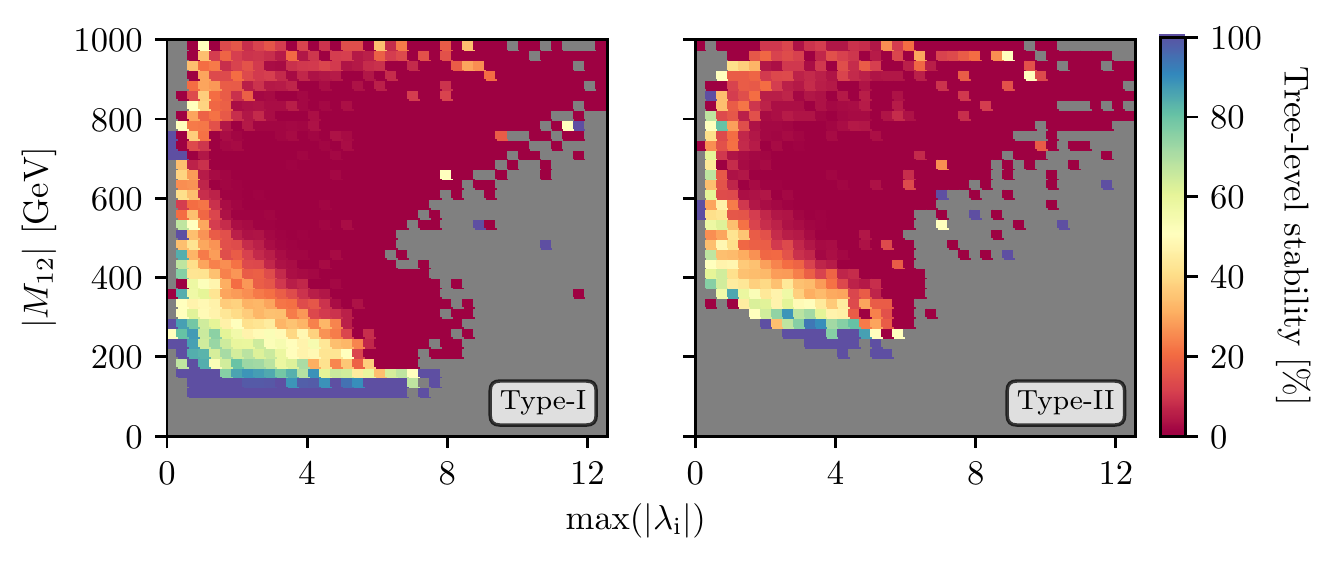}
\caption{Per-bin percentage of points that would have survived the one-loop vacuum stability constraints but would not have passed the tree-level conditions, as a function of $|M_{12}|$ and max$(|\lambda_i|)$, evaluated at the electroweak scale. The left plot shows the case of type-I Yukawas and the right one a Yukawa texture of type-II. }
\label{fig:vac_stab_tree_comparison}
\end{figure*}

Finally we comment on the conditions for electroweak vacuum stability. As discussed earlier, we use the one-loop effective potential in order to find all extrema {in the vicinity of the tree-level extrema} {checking} whether there {exists} a deeper global minimum. We only keep points which feature a stable desired electroweak vacuum configuration. The resulting constraints are in general different from the usual tree-level vacuum stability conditions, see Ref.~\cite{Staub:2017ktc} for more details. In \cref{fig:vac_stab_tree_comparison}, we show the fraction of parameter points in each bin which passed the one-loop constraints {which} would also have passed the tree-level vacuum stability conditions. For calculating the tree-level constraints, we again use the tree-level couplings obtained by \cref{eq:THDM:couplings_relations_1,eq:THDM:couplings_relations_2,eq:THDM:couplings_relations_3,eq:THDM:couplings_relations_4,eq:THDM:couplings_relations_5} and calculate the tree-level potential. 

In accordance with Ref.~\cite{Staub:2017ktc}, we find that large regions of parameter space which feature a perfectly fine EWSB global minimum at the one-loop order would have been regarded unstable by the tree-level checks -- meaning that these regions are resurrected by the radiative corrections. While for small $|M_{12}|$, the tree-level conditions would have allowed almost all of the parameter points, it is clearly seen that for larger values of $\mathcal{O}(\SI{400}{\GeV})$ and higher, far less than half of the points would have been {considered} allowed when applying the conventional checks. Interestingly, $M_{12}$ is the most decisive factor in the change of tree-level forbidden to loop-level allowed. The size of the quartic couplings instead plays an important though inferior role. In particular, for $|M_{12}|\gtrsim \SI{600}{\GeV}$ and max$(|\lambda_i|)_{\rm EWSB}\gtrsim 5$ (which is of course exactly the region where the scalar loop corrections are large and therefore also the corrections to the potential), virtually all the parameter space would be ruled out by the tree-level checks -- but not so once the radiative corrections are taken into account. The reason for this behaviour can be found in the size of the scalar loop corrections in this region: as discussed earlier, $M_{12}$ drives the (positive) loop corrections to the lightest Higgs mass. As a result, in an \MS scheme, we often need  {\emph{large negative}} $\lambda_{1,2}$ in order to obtain the correct Higgs mass at two-loop order. The area in the figure where almost none of the allowed points would have been allowed at the tree level corresponds to exactly this situation. As a matter of fact, although the threshold corrections drive $\lambda_{1,2}$ to quite large negative numbers, their tree-level -- or on-shell -- equivalents are usually also negative. However, since negative $\lambda_{1,2}$ lead to field directions which are unbounded from below at tree-level \cite{Deshpande:1977rw}, the tree-level calculation would result in the statement that these points are excluded. At the loop level, however, the situation is different as the large loop corrections can lift the potential in these unbounded-from-below directions and therefore stabilize the vacuum \cite{Staub:2017ktc}. Lastly note that the lack of parameter points on the right-hand side (displaying the THDM-II case) for both low $|M_{12}|$ and $|\lambda_i|$ is again due to the stronger cuts on $m_{H^\pm}$ which require either of both to be large in order to produce the large masses needed.

% --------------------------------------------
\section{Summary and Conclusions}
\label{sec:summary}
% --------------------------------------------

In this paper, we have studied generic predictions from UV completions of THDMs. We have not specified the particular UV-complete model but rather investigated  the low-energy consequences of general boundary conditions at a particular matching scale, i.e. leaving the THDM parameters arbitrary at this scale. By the use of the two-loop renormalisation group equations, those parameters were then evolved down to the electroweak scale where we also applied the two-loop threshold corrections for the Higgs mass. All obtained spectra have then been confronted with the current experimental constraints as well as the vacuum stability considerations. We further demanded perturbativity and perturbative unitarity of the theory everywhere between the TeV and the matching scale.  We have seen correlations between the matching scale and the mass splitting  $\Delta M$ in the heavy Higgs sector at the electroweak scale. As a generic feature, we find that large matching scales near the Planck scale would predict very small  $\Delta M$ independent of the size of the quartic couplings at the scale. If, in turn, this splitting should be of the order of several hundreds of GeV, this would point to very large couplings at a matching scale not much larger than the TeV scale probed so far at experiments, placing serve constraints on the possibility of realising electroweak baryogenesis in THDMs.

We have highlighted the importance of the loop corrections to the Higgs mass which need to be taken into account for reliable predictions. Likewise, we have shown that an examination of the stability of the electroweak vacuum needs to be done beyond tree level -- or else we would wrongly consider many perfectly-allowed regions of parameter space as ruled out.

\section*{Acknowledgements}
We thank Johannes Braathen and Mark D. Goodsell for useful discussions and the LPTHE in Paris for hospitality. MEK is supported by the DFG Research Unit 2239 ``New Physics at the LHC''. TO has received funding from the German Research Foundation (DFG) under Grant Nos.\ EXC-1098, FOR~2239 and GRK~1581, and from the European Research Council (ERC) under the European Union's Horizon 2020 research and innovation programme (grant agreement No.\ 637506, ``$\nu$Directions''). TO would also like to thank the CERN Theoretical Physics Department for hospitality and support. FS is supported by the ERC Recognition Award ERC-RA-0008 of the Helmholtz Association. \\

\appendix

% --------------------------------------------
\section{Calculation of the mass spectrum at the low scale}
\label{subsec:mass+spectrum}
% --------------------------------------------

The quartic couplings are not free parameters but predicted at the matching scale between the THDM and its UV completion. 
Therefore, it is necessary to treat them as \MS parameters and to perform a calculation of the scalar masses and mixings including the higher order corrections. In practice, we perform 
the following steps:
\begin{enumerate}
 \item The running couplings $\lambda_i(Q)$ and $M_{12}(Q)$ at the scale $Q=m_t$ are taken as input, while the SM parameters are evolved to this scale including all known SM corrections, i.e. three-loop running and two-loop matching for $g_3$ and $Y_t$.  
 \item The running VEVs $v_1(Q)$ and $v_2(Q)$ are calculated from:
 \begin{align}
  v_1 =& \frac{v(Q)}{\sqrt{1+t^2_\beta}}\,, \\
  v_2 =& \frac{t_\beta v(Q)}{\sqrt{1+t^2_\beta}}\,,  
 \end{align}
 where $v(Q)$ is the running VEV and $\tan\beta$ is taken as input which is also defined at $Q=m_t$. The running VEV is calculated from the gauge couplings and the \MS vector boson masses 
 \begin{equation}
 M_V^{\MS}(Q) = \sqrt{M_V^2 + \Pi^T_{VV}(M_V^2)}\,,
 \end{equation}
 where $M_V$ is the pole mass and $\Pi^T_{VV}(M_V^2)$ the \MS self energy calculated at the scale $Q$ with external momentum $M_V$.
 \item The tree-level tadpole equations are solved to obtain $m_1^2$ and $m_2^2$:
 \begin{align}
 T_1 \equiv \frac{\partial V}{\partial \phi_1}\Big|_{\phi_1 = v_1} &= v_1 m_1^2 + \frac{1}{4} \Big[4 v_2 M_{12}^2 + 4 \lambda_1 v_{1}^{3} \\
 &\quad+  v_1 v_{2}^{2} \Big(2 \Big(\lambda_3 + \lambda_4\Big) + 2 \lambda_5\Big)\Big]= 0\,, \notag\\
T_2 \equiv \frac{\partial V}{\partial \phi_2}\Big|_{\phi_2 = v_2} &=  v_2 m_2^2 + \frac{1}{4} \Big[4 v_1 M_{12}^2 + 4 \lambda_2 v_{2}^{3} \\
&\quad + v_2 v_{1}^{2} \Big(2 \Big(\lambda_3 + \lambda_4\Big) + 2\lambda_5\Big)\Big] = 0\,. \notag
\end{align}
\begin{widetext}
\item The tree-level masses are calculated by diagonalising the mass matrices
\begin{align} 
m^2_{h} &= \left( 
\begin{array}{cc}
\frac{1}{2} \Big(6 \lambda_1 v_{1}^{2}  + v_{2}^{2} \Big(\lambda_3 + \lambda_4 + {\lambda_5\Big)}\Big) + m^2_1 &\frac{1}{2} v_1 v_2 \Big(2 \Big(\lambda_3 + \lambda_4\Big) + 2 {\lambda_5} \Big) + {M_{12}^2}\\ 
\cdot  &\frac{1}{2} \Big(6 \lambda_2 v_{2}^{2}  + v_{1}^{2} \Big(\lambda_3 + \lambda_4 + {\lambda_5}\Big)\Big) + m^2_2\end{array} 
\right) \\
m^2_{A^0} &= \left( 
\begin{array}{cc}
\frac{1}{2} \Big(2 \lambda_1 v_{1}^{2}  + v_{2}^{2} \Big( \lambda_3 + \lambda_4- {\lambda_5} \Big)\Big) + m^2_1 &v_1 v_2 {\lambda_5}  + {M_{12}^2}\\ 
\cdot &\frac{1}{2} \Big(2 \lambda_2 v_{2}^{2}  + v_{1}^{2} \Big( \lambda_3 + \lambda_4 - {\lambda_5} \Big)\Big) + m^2_2\end{array} 
\right) +  \xi_{Z}M^2_Z \\
m^2_{H^-} &= \left( 
\begin{array}{cc}
\frac{1}{2} \lambda_3 v_{2}^{2}  + \lambda_1 v_{1}^{2}  + m^2_1 &\frac{1}{2} \Big(\lambda_4 + \lambda_5\Big)v_1 v_2  + 
%m_{12}^*\\ 
M_{12}^{2\,*}\\
\cdot &\frac{1}{2} \lambda_3 v_{1}^{2}  + \lambda_2 v_{2}^{2}  + m^2_2\end{array} 
\right) +  \xi_{W^-}M^2_{W^-} 
 \end{align} 
 \end{widetext}
\item The one- and two-loop corrections $\delta t_i$ to the tadpoles are calculated. The imposed renormalisation conditions are:
 \begin{equation}
  T_i + \delta t_i = 0 \,,\hspace{1cm}i=1,2\,,
 \end{equation}
 which cause shifts in the Lagrangian parameters $m_i^2$:
 \begin{equation}
 \label{eq:shiftsm2}
  m_i^2 \to m_i^2 + \delta m_i^2 \,,\hspace{1cm}i=1,2 \,.
 \end{equation}
\item The one- and two-loop self-energies for real scalars are calculated for external gauge eigenstates.  At the one-loop level, the full dependence on the external momenta is included, while at two-loop, the approximation $p^2=0$ as well as  the gauge-less limit, i.e. $g_1=g_2=0$, is used.  The pole masses are the eigenvalues of the loop-corrected mass matrix calculated as
 \begin{equation}
  M_{\phi}^{(2L)}(p^2) = \tilde{M}^{(2L)}_{\phi} - \Pi_{\phi}(p^2)^{(1L)} -   \Pi_{\phi}(0)^{(2L)}\,.
 \end{equation}
Here, $\tilde{M}_{\phi}$ is the tree-level mass matrix including the shifts of \cref{eq:shiftsm2}. Particular care is needed for the two-loop corrections because we work in the 
gaugeless limit where the Goldstone bosons are massless. Those cause IR divergences in the two-loop integrals. In order to avoid this so called `Goldstone bosons catastrophe', we use the approach
presented in Refs.~\cite{Braathen:2016cqe,Braathen:2017izn}.

For charged scalars, the scalar masses are available at the one-loop level,
\begin{equation}
 M_{\phi}^{(1L)}(p^2) = \tilde{M}^{(1L)}_{\phi} - \Pi_{\phi}(p^2)^{(1L)} \,.
\end{equation}
The calculation of the one-loop self-energies in both cases is done iteratively for each eigenvalue $i$ until the on-shell condition
\begin{equation}
  \left[\text{eig} M_{\phi}^{(n)}(p^2=m^2_{\phi_i})\right]_i \equiv m_{\phi_i}^2
\end{equation}
is fulfilled. 
\end{enumerate}

\bibliographystyle{JHEP}
\bibliography{literature}

\end{document}